\documentclass[12pt,aps,floats]{revtex4}
\usepackage{blindtext}
\usepackage{graphicx}
\usepackage{color}
\usepackage{amsmath}  
\usepackage{amssymb}
\usepackage{amsfonts} 
\usepackage{graphicx} 
\usepackage{xcolor}

\begin{document}
\title{The Bell theorem revisited: geometric phases in gauge theories}
\author{David H. Oaknin}
\affiliation{Rafael Ltd, IL-31021 Haifa, Israel, \\
e-mail: { d1306av@gmail.com}}

\begin{abstract}
The Bell theorem stands as an insuperable roadblock in the path to a very desired intuitive solution of the EPR paradox and, hence, it lies at the core of the current lack of a clear interpretation of the quantum formalism. The theorem states through an experimentally testable inequality that the predictions of quantum mechanics for the Bell polarization states of two entangled particles cannot be reproduced by any statistical model of hidden variables that shares certain intuitive features. In this paper we show, however, that the proof of the Bell theorem involves a subtle, though crucial, assumption that is not required by fundamental physical principles and, hence, it is not necessarily fulfilled in the experimental setup that tests the inequality. Indeed, this assumption can neither be properly implemented within the standard framework of quantum mechanics. Namely, the proof of the theorem assumes that there exists a preferred absolute frame of reference, supposedly provided by the lab, which enables to compare the orientation of the polarization measurement devices for successive realizations of the experiment and, hence, to define jointly their response functions over the space of hypothetical hidden configurations for all their possible alternative settings. We notice, however, that only the relative orientation between the two measurement devices in every single realization of the experiment is a properly defined physical degree of freedom, while their global rigid orientation is a spurious gauge degree of freedom. Hence, the preferred frame of reference required by the proof of the Bell theorem does not necessarily exist. In fact, it cannot exist in models in which the gauge symmetry of the experimental setup under global rigid rotations of the two detectors is spontaneously broken by the hidden configurations of the pair of entangled particles and a non-zero geometric phase appears under some cyclic gauge symmetry transformations. Following this observation, we build an explicitly local model of hidden variables that reproduces the predictions of quantum mechanics for the Bell states.
\end{abstract}


\maketitle

\newpage
\section{Introduction}

The Bell theorem is one of the fundamental theorems upon which relies the widespread belief that quantum mechanics is the ultimate mathematical framework within which the hypothetical final theory of the fundamental building blocks of Nature and their interactions should be formulated. The theorem states through an experimentally testable inequality (the Bell inequality) that statistical models of hidden variables that share certain intuitive features cannot reproduce the predictions of quantum mechanics for the entangled polarization states of two particles (the Bell states) \cite{Bell,Fine}. These predictions have been confirmed beyond doubt by very carefully designed experiments \cite{Hansen,Aspect,Gisin,Weihs,Rowe,Giustina,Christensen,Giustina2,Shalm,ScienceNews}.

In these experiments a source emits pairs of particles whose polarizations are arranged in a Bell entangled state:
\begin{equation}
\label{Bell_state}
| \Psi_{\Phi} \rangle = \frac{1}{\sqrt{2}} \left(|\uparrow \rangle^{(A)} \ |\downarrow\rangle^{(B)} - e^{i \Phi} \ |\downarrow \rangle^{(A)} \ |\uparrow\rangle^{(B)}\right), 
\end{equation}
where $\left\{|\uparrow\rangle, \ |\downarrow\rangle\right\}^{(A,B)}$ are eigenstates of Pauli operators $\sigma_Z^{(A,B)}$ along locally defined Z-axes for each one of the two particles. The two emitted particles travel off the source in opposite directions towards two widely separated detectors, which test their polarizations. The orientation of each one of the detectors can be freely and independently set along any arbitrary direction in the XY-plane perpendicular to the locally defined Z-axis.  Upon detection each particle causes a binary response of its detector, either $+1$ or $-1$. Thus, each pair of entangled particles produces an outcome in the space of possible events ${\cal P} \equiv \left\{(-1,-1), (-1,+1), (+1,-1), (+1,+1)\right\}$. We refer to each detected pair as a single realization of the experiment.

Quantum mechanics predicts that the statistical correlation between the binary outcomes of the two detectors in a long sequence of realizations of the experiment is given by:
\begin{equation}
\label{correlation}
E(\Delta,\Phi) = -\cos(\Delta-\Phi),
\end{equation} 
where $\Delta$ is the relative angle between the orientations of the two detectors. In particular, when $\Delta-\Phi=0$ we get that  $E=-1$, so that all outcomes in the sequence must be either $\left(-1, +1\right)$ or $\left(+1, -1\right)$.

The Bell theorem states that prediction (\ref{correlation}) cannot be reproduced by any model of hidden variables that shares certain intuitive features. In particular, the CHSH version of the theorem states that for the said generic models of hidden variables the following inequality is fulfilled for any set of values $\left(\Delta_1, \Delta_2, \delta\right)$ \cite{CHSH}: 
\begin{eqnarray}
\label{CHSH_original}
\left|E(\Delta_1) + E(\Delta_2) +  E(\Delta_1-\delta) - E(\Delta_2-\delta)\right| \le 2.  \ \  \ \  \ \ 
\end{eqnarray}   
On the other hand, according to quantum mechanics the magnitude in the left hand side of the inequality reaches a maximum value of $2\sqrt{2}$, known as Tsirelson's bound \cite{Tsirelson}, for certain values of $\Delta_1$, $\Delta_2$ and $\delta$ - {\it e.g}, $\Delta_1=-\Delta_2=\frac{1}{2}\delta=\frac{\pi}{4}$. As it was noted above, carefully designed experiments have confirmed that the CHSH inequality is violated according to the predictions of quantum mechanics and, therefore, have ruled out all the generic models of hidden variables constrained by the Bell inequality (\ref{CHSH_original}). 

In this paper we show, however, that the generic models of hidden variables constrained by the Bell theorem all share a subtle crucial feature that is not necessarily fulfilled in the actual experimental tests of the Bell inequality. Indeed, the considered feature cannot be derived from fundamental physical principles and may even be at odds with the fundamental principle of relativity. Moreover, this feature neither can be properly implemented within the standard framework of quantum mechanics. We follow this observation to explicitly build a local model of hidden variables that does not share the disputed feature and, thus, it is capable to reproduce the predictions of quantum mechanics for the Bell polarization states of two entangled particles. 

Our model puts forward for consideration the possibility that quantum mechanics might not be the ultimate mathematical framework of fundamental physics. In fact, it is interesting to notice that the way how our model solves the apparent 'non-locality' associated to entanglement in the standard quantum formalism is very similar to the way how General Relativity solves the 'non-locality' of Newton's theory of gravitation: in our model quantum entanglement is the result of a curved metric in the space in which the hypothetical hidden variables live.
\

\section{Outline}

Any local statistical model of hidden variables that aims to describe the Bell experiment consists of some space ${\cal S}$ of possible hidden configurations for the pair of entangled particles, labelled here as $\lambda \in {\cal S}$, together with a well-defined (density of) probability $\rho(\lambda)$ for each one of them to occur in every single realization. The model must also specify well-defined binary functions $s^{(A)}_{\Omega_A}(\lambda) = \pm 1$, $\ s^{(B)}_{\Omega_B}(\lambda) = \pm 1$ to describe the outcomes that would be obtained at detectors A and B when the pair of entangled particles occurs in the hidden configuration $\lambda \in {\cal S}$ and their polarizations are tested along directions $\Omega_A$ and $\Omega_B$, respectively.

The proof of the CHSH inequality (\ref{CHSH_original}) involves two well-defined possible orientations $\Omega_A$ and $\Omega'_A$ for the polarization test of particle A and two well-defined possible orientations $\Omega_B$ and $\Omega'_B$ for the polarization test of particle B, and assumes that the considered model of hidden variables assigns to each possible hidden configuration $\lambda \in {\cal S}$ a 4-tuple of binary values $\left(s^{(A)}_{\Omega_A}(\lambda), s^{(A)}_{\Omega'_A}(\lambda), s^{(B)}_{\Omega_B}(\lambda), s^{(B)}_{\Omega'_B}(\lambda)\right) \in \left\{-1,+1\right\}^4$ to describe the outcomes that would be obtained in each one of the two detectors in case that it would be set along each one of its two available orientations. Hence, it is straightforward to obtain that for any $\lambda \in {\cal S}$,
\begin{eqnarray}
\label{CHSH_proof}
s^{(A)}_{\Omega_A}(\lambda)\cdot \left(s^{(B)}_{\Omega_B}(\lambda) + s^{(B)}_{\Omega'_B}(\lambda)\right) 
+ \ s^{(A)}_{\Omega'_A}(\lambda)\cdot \left(s^{(B)}_{\Omega_B}(\lambda) - s^{(B)}_{\Omega'_B}(\lambda)\right)=\pm2,
\end{eqnarray}
since the first term is non-zero only when $s^{(B)}_{\Omega_B}(\lambda)$ and $ s^{(B)}_{\Omega'_B}(\lambda)$ have the same sign, while the second term is non-zero only when they have opposite signs. The CHSH inequality (\ref{CHSH_original}) is then obtained by averaging (\ref{CHSH_proof}) over the whole space ${\cal S}$ of all possible hidden configurations, since
\begin{eqnarray}
\label{CHSH_integral}
\left|\int d\lambda \ \rho(\lambda) \ \left\{s^{(A)}_{\Omega_A}(\lambda)\cdot \left(s^{(B)}_{\Omega_B}(\lambda) + s^{(B)}_{\Omega'_B}(\lambda)\right) 
+ \ s^{(A)}_{\Omega'_A}(\lambda)\cdot \left(s^{(B)}_{\Omega_B}(\lambda) - s^{(B)}_{\Omega'_B}(\lambda)\right)\right\}\right| \le 2,
\end{eqnarray}
while 
\begin{eqnarray}
\int d\lambda \ \rho(\lambda) \ s^{(A)}_{\Omega_A}(\lambda) \cdot s^{(B)}_{\Omega_B}(\lambda) & = & E(\Delta_1), \\
\int d\lambda \ \rho(\lambda) \ s^{(A)}_{\Omega_A}(\lambda) \cdot s^{(B)}_{\Omega'_B}(\lambda) & = & E(\Delta_2), \\
\int d\lambda \ \rho(\lambda) \ s^{(A)}_{\Omega'_A}(\lambda) \cdot s^{(B)}_{\Omega_B}(\lambda) & = & E(\Delta_1-\delta), \\
\int d\lambda \ \rho(\lambda) \ s^{(A)}_{\Omega'_A}(\lambda) \cdot s^{(B)}_{\Omega'_B}(\lambda) & = & E(\Delta_2-\delta).
\end{eqnarray}  
\
In this argument the orientations $\Omega_A$, $\Omega'_A$, $\Omega_B$ and $\Omega'_B$ seem to be fixed with respect to some external frame of reference supposedly provided by the labs. Nonetheless, the data collected in such an experimental setup could be alternatively analyzed within frames of reference aligned, for example, with the magnetic axis of the Sun or the rotational axis of the Galaxy, with respect to which the orientations of the detectors for different realizations of the experiment are not fixed anymore. Obviously, the conclusions of the analysis must remain the same, independently of the chosen lab frame. Indeed, the proof of the CHSH inequality actually requires only three well-defined angles, $\Delta_1 \equiv \angle (\Omega_B, \Omega_A)$, $\Delta_2 \equiv \angle (\Omega'_B, \Omega_A)$ and $\delta \equiv \angle (\Omega'_A, \Omega_A)$, which correspond, respectively, to the relative orientations of $\Omega_B$, $\Omega'_B$ and $\Omega'_A$ with respect to $\Omega_A$, which serves as a reference direction. The reference direction $\Omega_A$ serves also to define the hidden configuration $\lambda \in {\cal S}$ of the pair of entangled particles in every single realization of the experiment, since the description of a physical state must necessarily be done with respect to a reference frame.  Otherwise, the orientation with respect to any external lab frame, either the optical table or the stars in the sky, of this reference direction $\Omega_A$ at different single realizations of the Bell experiment is absolutely irrelevant: it is an spurious gauge degree of freedom, which can be set to zero (see Fig. 1). 

The proof of the CHSH inequality, thus, seems straightforward and unovaidable. Nonetheless, the main claim of this paper is that this proof, as well as the proofs of all other versions of the Bell inequality, involve a subtle, though crucial, implicit assumption that cannot be derived from fundamental physical principles and, indeed, it might not be fulfilled in the actual experimental setup that tests the inequality. Namely, in each realization of a Bell experiment the polarization of each one of the two entangled particles is tested along a single direction. Hence, the relative orientation $\Delta$ of the two measurement devices in each single realization of the experiment is a properly defined physical magnitude, which can be set to values $\Delta_1$, $\Delta_2$ or any other desired value. On the other hand, the definition of the angle $\delta$ that appears in the proof of the CHSH inequality requires a comparison of the global rigid orientation of the measurement devices for different realizations of the Bell experiment and, thus, it requires the existence of an absolute preferred frame of reference with respect to which this global orientation could be defined. Otherwise, we could choose the orientation of, say,  detector A as the reference direction for every single realization of the experiment and define the orientation of the other detector with respect to it,  in which case the proof of the Bell theorem does not necessarily hold as we shall show later. Obviously, such an absolute preferred frame of reference would not be needed if the polarization of each one of the two entangled particles could be tested along two different directions at once in every single realization of the experiment, but this is certainly not the case. Similar concerns regarding the way how different settings of the detectors are compared within the framework of the Bell theorem and the crucial role that this comparison plays in the proof of the inequality are also raised by K. Hess in \cite{Hess, Hess1}, and much earlier in a different but related context in \cite{Hess2,Hess3,Hess4}.

\begin{figure}
\centering
\vspace{0.5in}
\hspace{-0.15in}
\includegraphics[height=10.0cm]{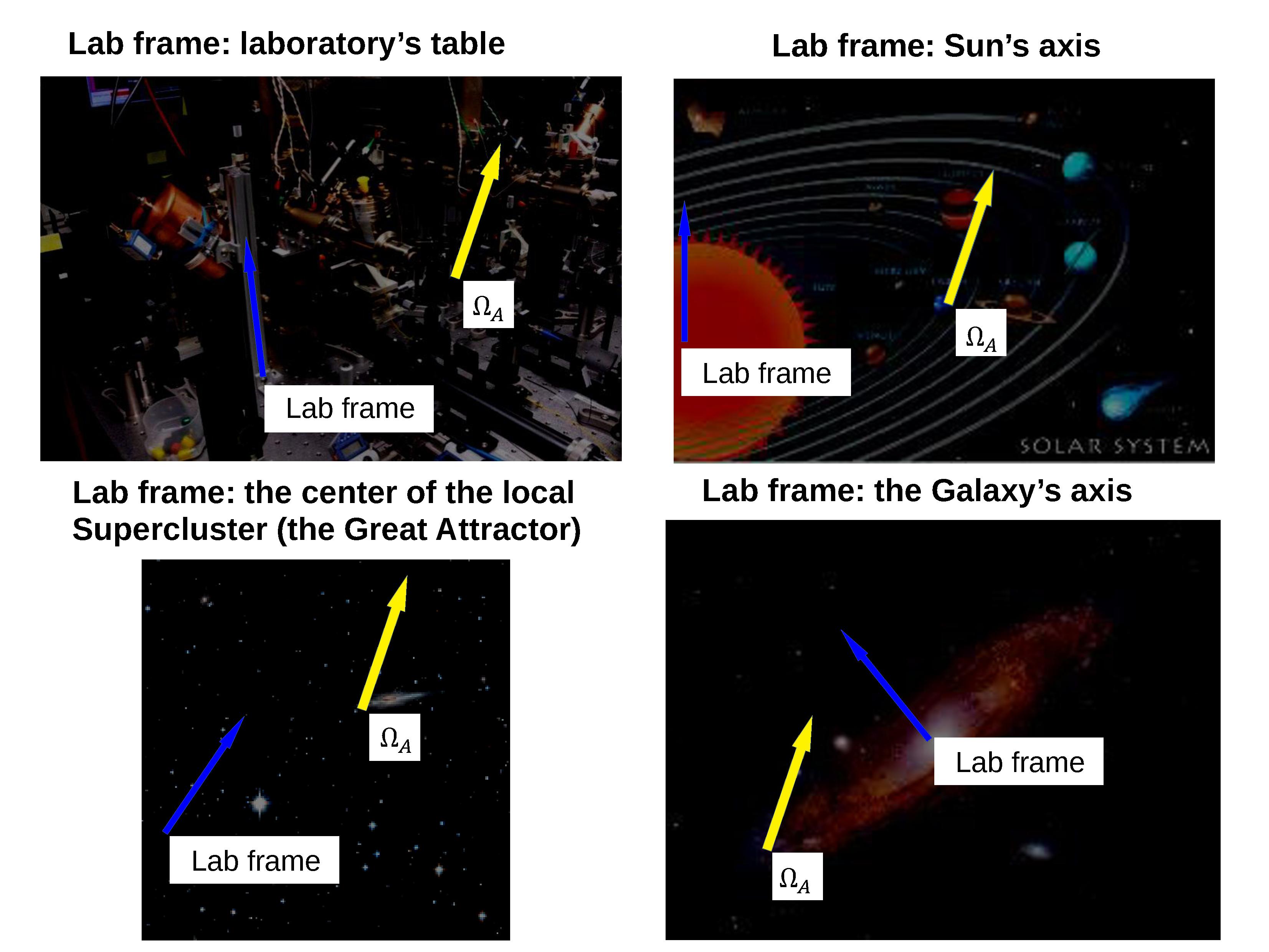}
\fontsize{10}{0}
\selectfont\put(-197.,90.){\makebox(0,0)[t]{\textcolor[rgb]{0,0,0}{}}}
\fontsize{10}{0}
\selectfont\put(-245.,187.){\rotatebox{90}{\makebox(0,0)[b]{\textcolor[rgb]{0,0,0}{}}}}
\vspace{-0.0in}
\caption{The orientation of the reference direction $\Omega_A$ with respect to the chosen lab frame is a spurious gauge degree of freedom.}
\vspace{0.5in}
\end{figure}

The said preferred frame of reference needed to prove the Bell theorem is supposedly provided by the lab. However, the conditions that a reference frame must fulfill in order to qualify as a preferred absolute frame are far from obvious and, in any case, its existence is an overbold assumption whose fulfillment has never been explored neither theoretically or experimentally. In fact, the existence of an absolute preferred frame of reference would be clearly at odds with Galileo's principle of relativity. Moreover, it is straightforward to show that this assumption cannot be properly implemented within the standard framework of quantum mechanics either. The argument goes as follows. The Bell state (\ref{Bell_state}) that describes the pair of entangled particles is defined in terms of the bases $\{|\uparrow \rangle, \ |\downarrow \rangle\}^{(A,B)}$ of eigenstates of the Pauli operators $\sigma_Z^{(A,B)}$ along locally defined Z-axes for each one of the particles. Since these eiegenstates are defined up to a global phase, the phase $\Phi$ in expression (\ref{Bell_state}) cannot be properly defined with respect to a lab frame. In order to properly define it we need to choose an arbitrary setting of the two detectors that test the polarizations of the pair of entangled particles as a reference. This reference setting defines {\it parallel} directions along the XY-planes at the sites where each one of the two particles are detected. Then, the phase $\Phi$ of the entangled state  (\ref{Bell_state}) can be properly defined  with respect to this reference setting with the help of the measured correlations between the outcomes of the two detectors, $E=-\cos(\Phi)$. Furthermore, we can use this reference setting to properly define a relative rotation $\Delta$ of the orientations of the two measurement devices. On the other hand, since we must use an arbitrary setting of the detectors as a reference, their absolute orientation is an unphysical gauge degree of freedom (see Fig. 2). In summary, in order to describe the setting of the measurement devices in a Bell experiment within the standard framework of quantum mechanics we need to specify both $\Phi$ and $\Delta$ with respect to some otherwise arbitrary reference setting of the detectors. Nonetheless, only their difference $\Delta - \Phi$ is independent of the chosen reference setting and, hence, the correlation between the outcomes of the two devices can only depend on this difference (\ref{correlation}).

In the absence of an absolute preferred frame of reference the global rigid orientation of the two detectors is, as we have already noticed before, an spurious (unphysical) gauge degree of freedom and, hence, the proof of the CHSH inequality (as well as of all other versions of the Bell inequality) holds only for models in which the considered hidden configurations are symmetrically invariant under a rigid rotation of the two measuring devices. On the other hand, we shall show below that the proof of the inequality does not necessarily hold when this symmetry is (spontaneously) broken by the hidden configuration of the entangled particles, since then a non-zero geometric phase may appear under cyclic gauge transformations. Indeed, the crucial role of the angle $\delta$ in the proof of the CHSH inequality is an obvious indication that  in order to violate it the gauge symmetry under a rigid rotation of the two detectors must be spontaneously broken. 

In fact, it is obvious from the correlation (\ref{correlation}) that the entanglement of the two particles explicitly breaks the symmetry of the system under a rotation of the relative orientation of the two detectors. Since a reference direction is needed for this symmetry to get broken, the gauge symmetry under a rigid rotation of the two detectors must be also spontaneously broken. From this perspective the phase $\Phi$ that appears in the description of the source (\ref{Bell_state}) seems to play the role of a Goldstone mode associated to the spontaneously broken gauge symmetry, that is, the phase $\Phi$ appears instead of the spurious gauge degree of freedom $\delta$ that would describe the global rigid orientation of the two detectors. Under these circumstances, it is not possible to compare different settings of the detectors with respect to an external lab frame of reference: they can only be compared with respect to a frame in which they all share the same preferred direction, {\it e.g.} the reference frame set by the orientation of one of the detectors. This requirement can be explained as follows.

In the proof of the CHSH inequality it is implicitly assumed, as we have already noticed above, that there exists a preferred frame of reference, which defines a set of coordinates $\lambda \in {\cal S}$ over the space ${\cal S}$ of all possible hidden configurations that can be used to describe the response function of each one of the two detectors in each one of its two available orientations (defined with respect to the said preferred frame). Above we denoted these response functions as $s^{(A)}_{\Omega_A}(\lambda), s^{(A)}_{\Omega'_A}(\lambda), s^{(B)}_{\Omega_B}(\lambda), s^{(B)}_{\Omega'_B}(\lambda)$. Nonetheless, in general, we should allow for each one of the two detectors to define its proper set of coordinates over the space ${\cal S}$. Thus, for a given setting of the detectors we shall denote as $\lambda_A$ and $\lambda_B$ the sets of coordinates associated to detector A and detector B, respectively, so that their responses would be given as $s(\lambda_A)$ and $s(\lambda_B)$ by some universal function $s(\cdot)$ of the locally defined coordinate of the hidden configuration. Since these two sets of coordinates parameterize the same space of hidden configurations ${\cal S}$ there must exist some invertible transformation that relates them:

\begin{eqnarray}
\label{transform}
\lambda_B & = - L(\lambda_A; \ \Delta-\Phi),
\end{eqnarray}
which may depend parametrically on the relative orientation $\Delta-\Phi$ between the two detectors. This transformation must fulfill the constraint

\begin{equation}
d\lambda_A \ \rho(\lambda_A) = d\lambda_B \ \rho(\lambda_B),
\end{equation}
in order to guarantee that the probability of every hidden configuration to occur remains invariant under a change of coordinates, while the (density of) probability $\rho(\cdot)$ is functionally invariant for both sets of coordinates. However, these constraints do not forbid the possibility that the set of coordinates accumulates a non-zero geometric phase $\alpha \neq 0$ through certain cyclic gauge transformations:

\begin{eqnarray}
\label{geometric_phase}
\left(-L_{{\bar \Delta}_2}\right)  \circ \left(- L_{{\bar \Delta}_2-{\bar \delta}}\right) \circ \left(- L_{{\bar \Delta}_1-{\bar \delta}}\right) \circ \left(-L_{{\bar \Delta}_1}\right) \neq \mathbb{I},
\end{eqnarray}
In such a case there does not exist a single set of coordinates that can be used to define the response functions of each one of the two detectors in its two available orientations (defined with respect to an external frame), as required by the proof of the inequality (\ref{CHSH_original}). Therefore, in order to compare the four different experiments involved in the CHSH inequality we must choose the orientation of one of the detectors as a reference direction, as we do below in (\ref{new_model}), so that they all may be described within a common set of coordinates. The appearence of a non-zero geometric phase under a cyclic transformation is a well-known phenomena in physical models involving gauge symmetries \cite{Wilczek} and, therefore, we should not rule out the possibility that it also occurs in models of hidden variables for the Bell states. The Bell theorem, however, cannot account for such models. 

Following these observations we were able to explicitly build a local model of hidden variables that reproduces the predictions of quantum mechanics for the Bell polarization states. In our model the hidden configurations of the pair of entangled particles are described by a {\it pointer}, which sets an arbitrarily oriented preferred direction and, thus, spontaneously breaks the symmetry of the setup under rigid rotations of the two detectors. As we have just noticed, and we shall show later on in further detail, in order to compare different realizations of the experiment within the framework of such a model we must choose a common reference direction, which can be either the orientation of the hidden configuration of the pair of entangled particles or, alternatively, the orientation of one of the detectors, say, detector A. Since the former may not be directly experimentally accessible, we are left only with the latter option. Thus, in such a model we only need to specify the binary values for $s(\lambda_A)$, $s(\lambda_B)$,  $s(\lambda'_B)$, $s(\lambda''_B)$ and $s(\lambda'''_B)$ for each possible hidden configuration $\lambda_A \in {\cal S}$ of the pair of entangled particles, where $\lambda_B=-L(\lambda_A; \Delta_1)$, $\lambda'_B=-L(\lambda_A; \Delta_2)$, $\lambda''_B=-L(\lambda_A; \Delta_1-\delta)$, $\lambda'''_B=-L(\lambda_A; \Delta_2-\delta)$. It is then straightforward to notice that the magnitude 
\begin{equation}
\label{new_model}
s(\lambda_A) \ \cdot \ \left(s(\lambda_B) \ + \ s(\lambda'_B) \ + \ s(\lambda''_B) \ - \ s(\lambda'''_B)\right),
\end{equation}
which comes instead of (\ref{CHSH_proof}), can take values out of the interval $\left[-2, 2\right]$. Hence, these models are not constrained by the CHSH inequality (\ref{CHSH_original}). A simplified version of these arguments is presented in Fig. 3 with the help of a toy model.

\begin{figure}
\centering
\vspace{0.25in}
\hspace{-0.15in}
\includegraphics[height=10.0cm]{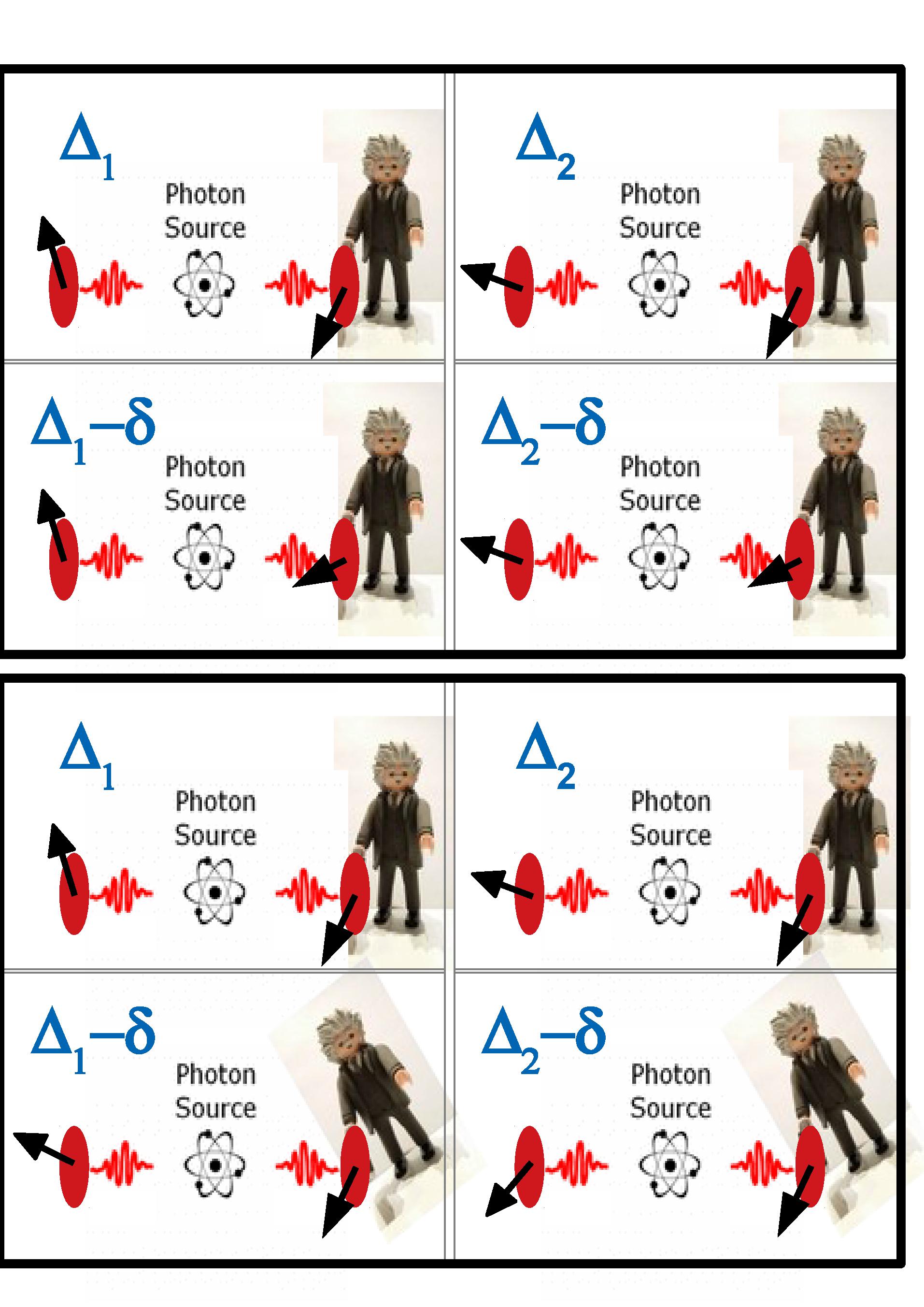}
\fontsize{10}{0}
\selectfont\put(-197.,90.){\makebox(0,0)[t]{\textcolor[rgb]{0,0,0}{}}}
\fontsize{10}{0}
\selectfont\put(-245.,187.){\rotatebox{90}{\makebox(0,0)[b]{\textcolor[rgb]{0,0,0}{}}}}
\vspace{-0.0in}
\caption{Two descriptions of the experimental setup required for testing the Bell inequality. In the description  above the lab frame is taken to be fixed, while in the description below the orientation of detector A is taken to be fixed. The relative angle between the two detectors is set at four possible values $\Delta_1$, $\Delta_2$, $\Delta_1-\delta$ and $\Delta_2-\delta$. When considering models in which the hypothetical hidden configurations of the pairs of entangled particles spontaneously break the symmetry under rigid rotations of the orientations of the two measurement devices, only the latter choice allows to properly compare the four different settings.}
\vspace{0.5in}
\end{figure}

\begin{figure}
\centering
\vspace{-1.0in}
\hspace{-0.15in}
\includegraphics[height=10.0cm]{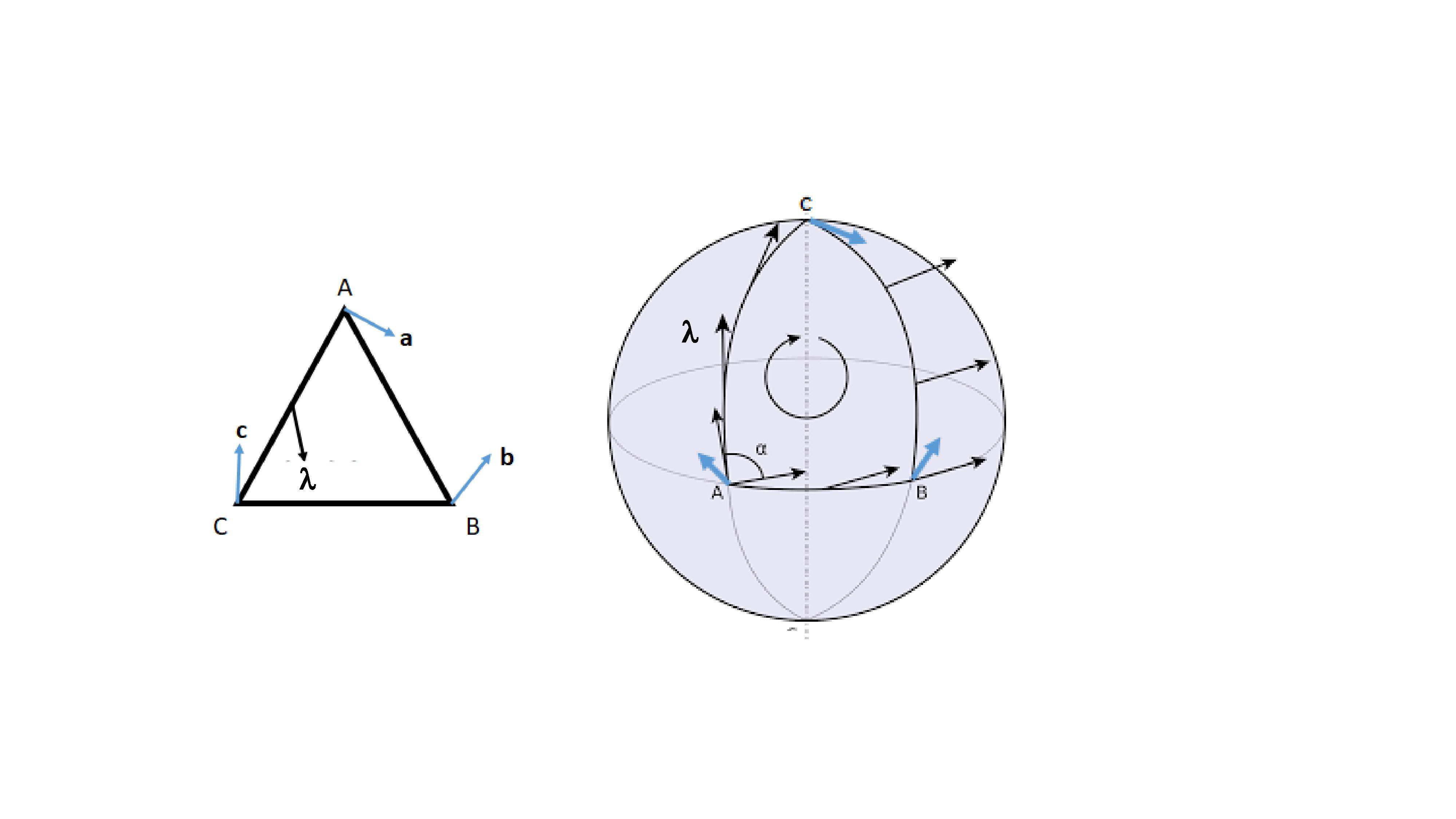}
\fontsize{10}{0}
\selectfont\put(-197.,90.){\makebox(0,0)[t]{\textcolor[rgb]{0,0,0}{}}}
\fontsize{10}{0}
\selectfont\put(-245.,187.){\rotatebox{90}{\makebox(0,0)[b]{\textcolor[rgb]{0,0,0}{}}}}
\vspace{-1.0in}
\caption{Two closely related, though intrinsically different, random games: the game on the left hand side is constrained by the Bell inequality, while the one on the right hand side is not necessarily constrained by the inequality. In both games we have reference unit vectors, labelled respectively as ${\vec a}$, ${\vec b}$ and ${\vec c}$, drawn at each one of the vertices, labelled as $A$, $B$ and $C$, of a triangle. In the game on the left the triangle is drawn on a plane surface and the reference unit vectors are contained within the plane, while in the game on the right the 'triangle' is defined on the surface of a sphere by segments of three great circles and the three reference unit vectors lay within the corresponding tangent planes. Two copies of a randomly oriented unit vector ${\vec \lambda}$ are generated at random at the center of one of the three segments of the triangle with density of probability $\rho({\vec \lambda})$, and detected, respectively, at the two detectors located at the ends of the segment. In the game on the left the vector ${\vec \lambda}$ is contained within the plane surface, while in the game on the right the vector ${\vec \lambda}$ is tangent to the sphere. The binary responses of the detectors are locally defined by parallely transporting the unit vector ${\vec \lambda}$ along the segment of the triangle to its end, and comparing its orientation to the orientation of the corresponding reference unit vector: $A({\vec a},{\vec \lambda})=\mbox{sign}({\vec a}\cdot{\vec \lambda})$, $B({\vec b},{\vec \lambda})=\mbox{sign}({\vec b}\cdot{\vec \lambda})$ , $C({\vec c},{\vec \lambda})=\mbox{sign}({\vec c}\cdot{\vec \lambda})$. It is then straighforward to prove the Bell inequality for the game on the left, since for any settings ${\vec a}, {\vec b}, {\vec c}$ and any random vector ${\vec \lambda}$ the following equality holds: 
$\left|A({\vec a},{\vec \lambda})\cdot B({\vec b},{\vec \lambda}) + A({\vec a},{\vec \lambda})\cdot C({\vec c},{\vec \lambda})\right| = 1 +  B({\vec b},{\vec \lambda})\cdot C({\vec c},{\vec \lambda})$. Therefore, after integrating over the whole space of possible hidden configurations: 
\newline
$\left|\int d{\vec \lambda} \ \rho({\vec \lambda}) \left[A({\vec a},{\vec \lambda})\cdot B({\vec b},{\vec \lambda}) + A({\vec a},{\vec \lambda})\cdot C({\vec c},{\vec \lambda})\right]\right| \le \int d{\vec \lambda} \ \rho({\vec \lambda}) \left|A({\vec a},{\vec \lambda})\cdot B({\vec b},{\vec \lambda}) + A({\vec a},{\vec \lambda})\cdot C({\vec c},{\vec \lambda})\right| = \int d{\vec \lambda} \ \rho({\vec \lambda}) \left|A({\vec a},{\vec \lambda})\cdot B({\vec b},{\vec \lambda}) + A({\vec a},{\vec \lambda})\cdot B({\vec b},{\vec \lambda})\cdot B({\vec b},{\vec \lambda})\cdot C({\vec c},{\vec \lambda})\right|= 1 +  \int d{\vec \lambda} \ \rho({\vec \lambda}) \ B({\vec b},{\vec \lambda})\cdot C({\vec c}, {\vec \lambda})$, and therefore, 
$\left|E^{A,B}({\vec a}, {\vec b}) + E^{A,C}({\vec a}, {\vec c})\right| \le 1 + E^{B,C}({\vec b},{\vec c})$. This proof, nonetheless, does not hold for the random game on the right hand side, since the orientation of a vector ${\vec \lambda}$ parallelly transported along the closed contour of the triangle $ABC$ gets rotated by a geometric phase $\alpha$ due to the curvature of the sphere. In fact, in the game on the right the three bipartite correlations are constrained by the inequality
$\left|E^{A,B}({\vec a}, {\vec b}) + E^{A,C}({\vec a}, {\vec c})\right| \le 1 + E^{B,C}({\vec b},{\hat {\cal R}}_{\alpha}{\vec c})$, where ${\hat {\cal R}}_{\alpha}{\vec c}$ denotes the rotation of vector ${\vec c}$ by an angle $\alpha$.}
\vspace{0.5in}
\end{figure}

\vspace{1.4in}
\hspace{1.00in}
\begin{tabular}{|r|r|r|r|r|}
\hline
$
\begin{array}{ccc}
\mbox{Outcome}\\
\mbox{Setting}
\end{array}
$&$
\begin{array}{ccc}
a&=&+1\\
b&=&+1
\end{array}
$&$
\begin{array}{ccc}
a&=&+1\\
b&=&-1
\end{array}
$&$
\begin{array}{ccc}
a&=&-1\\
b&=&+1
\end{array}
$&$
\begin{array}{ccc}
a&=&-1\\
b&=&-1
\end{array}
$\\
\hline
$
\begin{array}{ccc}
A&=&+1\\
B&=&+1
\end{array}
$&$
p_1 \ \ \ \ 
$&$
0 \ \ \ \ \ 
$&$
0 \ \ \ \ \ 
$&$
1 - p_1 \ \ 
$\\
\hline
$
\begin{array}{ccc}
A&=&+1\\
B&=&-1
\end{array}
$&$
p_2 \ \ \ \ 
$&$
0 \ \ \ \ \ 
$&$
0 \ \ \ \ \ 
$&$
1 - p_2 \ \ 
$\\
\hline
$
\begin{array}{ccc}
A&=&-1\\
B&=&+1
\end{array}
$&$
p_3 \ \ \ \ 
$&$
0 \ \ \ \ \ 
$&$
0 \ \ \ \ \ 
$&$
1 - p_3 \ \ 
$\\
\hline
$
\begin{array}{ccc}
A&=&-1\\
B&=&-1
\end{array}
$&$
0 \ \ \ \ \ 
$&$
p_4 \ \ \ \ 
$&$
1 - p_4 \ \
$&$
0 \ \ \ \ \  
$\\
\hline
\end{tabular}
\vspace{0.15in}
\\
Table 1. Conditional probabilities for a toy model with two binary inputs and two binary outcomes that cannot be reproduced by a realistic and local underlying theory \cite{PRboxes}. 
\vspace{0.2in}
\\

These arguments can be stated in more abstract terms as follows. Quantum predictions for the Bell experiment are commonly described as a set of conditional probabilities $p(a,b | A,B)$, where $a=\pm 1$ and $b=\pm 1$ are the two possible outcomes at each one of the two detectors and $A= \pm 1$ and $B =\pm 1$ describe two possible choices for the setting of each one of the two detectors. It is then proven that these conditional probabilities cannot be obtained in terms of a local model of hidden variables, defined by its configuration space $\lambda \in {\cal S}$, its density of probability $\rho(\lambda)$ and its local response functions $a=f(\lambda, A)$, $b=f(\lambda, B)$ \cite{Fine}. 
\\

This statement can be clearly illustrated with the help of the toy model described in Table~1 \cite{PRboxes}, where conditional probabilities for each one of the four possible results of an experiment with two binary outcomes $a,b=\pm1$ (columns) are given for each one of four possible settings, defined by two independent binary inputs $A,B=\pm1$ (rows). For these probabilities to be properly defined we require that $p_1, p_2, p_3, p_4 \in [0, 1]$. It can be readily checked that for each set of input values (rows) the sum of the probabilities for all possible results of the experiment (columns)  equals 1.  These conditional probabilities, however, cannot be obtained within the framework of an underlying local model of hidden variables: the conditional probabilities listed in the first three rows would imply $a = b$, that is, the outcomes of the two detectors in any of their four possible settings must have the same sign, which is obviously inconsistent with the conditional probabilities listed in the fourth row.

Nonetheless, it is straightforward to identify in this abstract reformulation of the Bell theorem the same unjustified implicit assumption that we have noticed above, namely, that there are two well-defined choices for the setting of each one of the detectors. We have noticed above that we can properly define and measure only the conditional probabilities $p(a,b | D)$, where $a=\pm 1$ and $b=\pm 1$ are, as before, the outcomes at each one of the two detectors and $D=1,2,3,4$ defines four possible relative orientations between them. We did notice also that quantum mechanics as well makes theoretical predictions only for these conditional probabilities $p(a,b | D)$. Under these looser constraints the Bell theorem does not necessarily hold.

Consider, for example, the toy model described in Table 2. The conditional probabilities are identical to those described in Table 1 for each one of the four possible results of the experiment, but the setting of the measurement devices is now described by a single parameter $D=1,2,3,4$. Each input value corresponds to a given relative orientation of the two devices. The new model simply states that when the devices are set at $D=1,2,3$ their outcomes must have the same sign, and when they are set at $D=4$ their outcomes must have opposite signs. Obviously, this latter model is not necessarily in contradiction with an underlying local model of hidden variables. \\

A straightforward proof of the inequalities that constraint the correlations that can be obtained in any model of hidden variables with two binary inputs and two binary outcomes is presented in \cite{Revzen} using only Boolean logic. The analysis relies on the observation that any such model makes a prediction for the correlations $\langle A B\rangle$, $\langle A B'\rangle$,  $\langle A' B\rangle$ and $\langle A' B'\rangle$, and also for the correlations $\langle A A'\rangle$ and $\langle B B'\rangle$ that would be obtained in the hypothetical case that the polarization of each one of the two entangled particles could be tested along two different orientations at once. It can be immediately noticed that these constraints do not hold for the model of hidden variables discussed in this paper, for which the correlations  $\langle A A'\rangle$ and $\langle B B'\rangle$ cannot be jointly bounded.

\vspace{0.3in}
\hspace{1.0in}
\begin{tabular}{|r|r|r|r|r|}
\hline
$
\begin{array}{ccc}
\mbox{Outcome}\\
\mbox{Setting}
\end{array}
$&$
\begin{array}{ccc}
a&=&+1\\
b&=&+1
\end{array}
$&$
\begin{array}{ccc}
a&=&+1\\
b&=&-1
\end{array}
$&$
\begin{array}{ccc}
a&=&-1\\
b&=&+1
\end{array}
$&$
\begin{array}{ccc}
a&=&-1\\
b&=&-1
\end{array}
$\\
\hline
$
\begin{array}{ccc}
D&=&1\\
\end{array}
$&$
p_1 \ \ \ \ 
$&$
0 \ \ \ \ \ 
$&$
0 \ \ \ \ \ 
$&$
1 - p_1 \ \ 
$\\
\hline
$
\begin{array}{ccc}
D&=&2\\
\end{array}
$&$
p_2 \ \ \ \ 
$&$
0 \ \ \ \ \ 
$&$
0 \ \ \ \ \ 
$&$
1 - p_2 \ \ 
$\\
\hline
$
\begin{array}{ccc}
D&=&3\\
\end{array}
$&$
p_3 \ \ \ \ 
$&$
0 \ \ \ \ \ 
$&$
0 \ \ \ \ \ 
$&$
1 - p_3 \ \ 
$\\
\hline
$
\begin{array}{ccc}
D&=&4\\
\end{array}
$&$
0 \ \ \ \ \ 
$&$
p_4 \ \ \ \ 
$&$
1 - p_4 \ \
$&$
0 \ \ \ \ \  
$\\
\hline
\end{tabular}
\vspace{0.15in}
\\
Table 2. Conditional probabilities for a toy model with a single input with four possible values and two binary outcomes. They can be reproduced by an underlying theory. 
\vspace{0.2in}
\\


\section{The model}

We shall now build an explicitly local statistical model of hidden variables that reproduces the predictions of quantum mechanics for the Bell states (\ref{Bell_state}) and, hence, it is not constrained by the Bell inequality (\ref{CHSH_original}). The fundamental ideas of the model were first discussed in \cite{david}. As we have already noticed above, the crux of the model is the spontaneous breaking of the gauge symmetry of the experimental setup under global rigid rotations of the orientation of the detectors. The symmetry is broken by the hidden configuration of the pair of entangled particles. Furthermore, we allow for a non-zero geometric phase (\ref{geometric_phase}) to accumulate through cyclic gauge transformations. Under these circumstances there does not exist an absolute preferred frame, other than the orientation of one of the detectors, to which we can refer in order to compare different realizations of the experiment (see Fig. 3). 

\begin{figure}
\centering
\vspace{-1.0in}
\includegraphics[height=13cm]{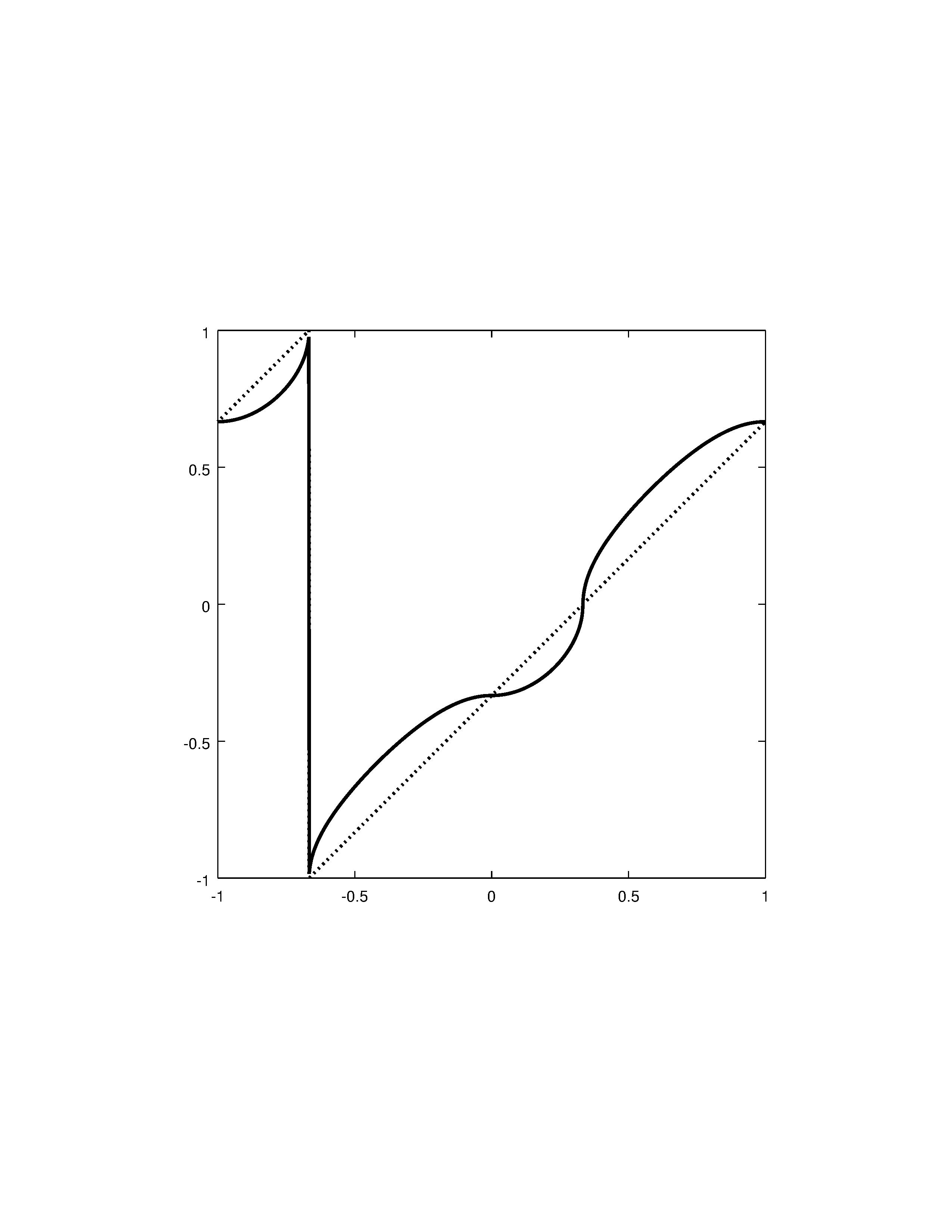}
\fontsize{10}{0}
\selectfont\put(-137.,90.){\makebox(0,0)[t]{\textcolor[rgb]{0,0,0}{{\bf $\lambda/\pi$}}}}
\fontsize{10}{0}
\selectfont\put(-245.,187.){\rotatebox{90}{\makebox(0,0)[b]{\textcolor[rgb]{0,0,0}{{\bf $\lambda'/\pi$}}}}}
\vspace{-1.0in}
\caption{Plot of the transformation law $\lambda \rightarrow \lambda' = L(\lambda; \Delta)$ for $\Delta=\pi/3$ (solid line), compared to the corresponding linear transformation (dotted line).}
\end{figure}

The gauge symmetry is spontaneously broken because in the considered model the hidden configuration of the pair of entangled particles has a preferred direction randomly oriented over a unit circle ${\cal S}$ in the XY-plane. This orientation is carried by each one of the particles of the entangled pair. Each one of the two detectors defines over this circle ${\cal S}$ a frame of reference with its own set of associated coordinates, which we shall denote as $\lambda_A\in [-\pi, +\pi)$ for detector A and $\lambda_B \in [-\pi, +\pi)$ for detector B. Since the two sets of coordinates parameterize the same space ${\cal S}$, they must be related by some transformation law:

\begin{equation}
\label{Oaknin}
\lambda_B =-L(\lambda_A; \ \Delta - \Phi),
\end{equation}
where $\Delta$ is the relative angle between the two detectors and $\Phi$ is the phase that characterizes the source of entangled particles as defined above. This transformation law states that a hidden configuration whose preferred direction is oriented along an angle $\lambda_A$ with respect to detector A, it is oriented along an angle $\lambda_B$ with respect to detector B. 

The transformation law (\ref{Oaknin}) does not violate neither locality nor causality: it may well be a fundamental law of Nature. Indeed, the notions of locality and causality in special relativity stem from a similar relationship $v' = T(v; \ V)$ beween the velocities $v$ and $v'$ of a point particle with respect to two different inertial frames moving with relative velocity $V$. Moreover, (\ref{Oaknin}) is only a generalization of the euclidean linear relationship that states that in a flat space given two detectors whose orientations form an angle $\Delta$, then a pointer oriented along an angle $\omega$ with respect to one of them is oriented along an angle $\omega - \Delta$ with respect to the other detector.

In order to reproduce the predictions of quantum mechanics we define the transformation law (\ref{Oaknin}) as follows:

\begin{itemize}
\item If  ${\bar \Delta} \in [0, \pi)$, 
\begin{eqnarray}
\label{Oaknin_transformation}
\hspace{-0.15in}
L(\lambda; {\bar \Delta}) =  
\left\{
\begin{array}{c}
\hspace{-0.0in} q(\lambda-{\bar \Delta}) \cdot \mbox{arccos}\left(-\cos({\bar \Delta}) - \cos(\lambda) - 1 \right), \\ \hspace{0.88in} \mbox{if}  \hspace{0.1in} -\pi \hspace{0.16in} \le  \lambda < {\bar \Delta}-\pi, \\
\hspace{-0.0in} q(\lambda-{\bar \Delta}) \cdot \mbox{arccos}\left(+\cos({\bar \Delta}) + \cos(\lambda) - 1 \right), \\ \hspace{0.685in} \mbox{if}  \hspace{0.05in} {\bar \Delta}-\pi \hspace{0.08in} \le \lambda < \hspace{0.105in} 0, \\
\hspace{-0.0in} q(\lambda-{\bar \Delta}) \cdot \mbox{arccos}\left(+\cos({\bar \Delta}) - \cos(\lambda) + 1 \right), \\ \hspace{0.69in} \mbox{if}  \hspace{0.25in} 0 \hspace{0.18in} \le \lambda < \ {\bar \Delta}, \\
\hspace{-0.0in} q(\lambda-{\bar \Delta}) \cdot \mbox{arccos}\left(-\cos({\bar \Delta}) + \cos(\lambda) + 1 \right), \\ \hspace{0.72in} \mbox{if}  \hspace{0.21in} {\bar \Delta}  \hspace{0.19in} \le  \lambda  < +\pi, \\
\end{array}
\right.
\end{eqnarray}
\item If  ${\bar \Delta} \in [-\pi, 0)$, 
\begin{eqnarray}
\label{Oaknin_transformation_Inv}
\hspace{-0.15in}
L(\lambda; {\bar \Delta}) =  
\left\{
\begin{array}{c}
q(\lambda-{\bar \Delta}) \cdot \mbox{arccos}\left(-\cos({\bar \Delta}) + \cos(\lambda) + 1 \right), \\ \hspace{0.650in} \mbox{if}   \hspace{0.11in} -\pi \hspace{0.15in} \le \lambda < {\bar \Delta}, \\
q(\lambda-{\bar \Delta}) \cdot \mbox{arccos}\left(+\cos({\bar \Delta}) - \cos(\lambda) + 1 \right), \\ \hspace{0.66in} \mbox{if}   \hspace{0.30in} {\bar \Delta} \hspace{0.1in} \le \lambda < \hspace{0.05in} 0, \\
q(\lambda-{\bar \Delta}) \cdot \mbox{arccos}\left(+\cos({\bar \Delta}) + \cos(\lambda) - 1 \right), \\ \hspace{0.92in} \mbox{if}  \hspace{0.23in} 0 \hspace{0.22in} \le \lambda < {\bar \Delta} +\pi, \\
q(\lambda-{\bar \Delta}) \cdot \mbox{arccos}\left(-\cos({\bar \Delta}) - \cos(\lambda) - 1 \right), \\ \hspace{0.75in} \mbox{if}  \hspace{0.16in} {\bar \Delta} +\pi \hspace{0.00in} \le \lambda < +\pi, \\
\end{array}
\right.
\end{eqnarray}
\end{itemize}
where 
\begin{eqnarray*}
q(\lambda-{\bar \Delta}) = \mbox{sign}((\lambda - {\bar \Delta}) \mbox{mod} ([-\pi, \pi))),
\end{eqnarray*}
${\bar \Delta} = \Delta - \Phi$ and the function $y=\mbox{arccos}(x)$ is defined in its main branch, such that $y \in [0, \pi]$ while $x \in [-1, +1]$. In Fig. 4 the transformation $L(\lambda; {\bar \Delta})$ is graphically shown for the particular case ${\bar \Delta} = \pi/3$.  It is straightforward to check that the transformation law (\ref{Oaknin}) is strictly monotonic and fulfills the differential relationship
\begin{equation}
\label{free}
\left|d\left(\cos(\lambda_B)\right)\right| = d\lambda_B \left|\sin(\lambda_B)\right| = d\lambda_A \left|\sin(\lambda_A)\right| = \left|d\left(\cos(\lambda_A)\right)\right|,
\end{equation}
while the parameter ${\bar \Delta}$ plays the role of an the integration constant.

Locality is explicitly enforced in our model by requiring that the outcome of each one of the detectors in reponse to the hidden configuration of the pair of entangled particles depends only on its locally defined orientation, that is, $s(\lambda_A) = \pm 1$ for detector A and $s(\lambda_B) = \pm 1$ for detector B, where $\lambda_B$ and $\lambda_A$ are related by relationship (\ref{Oaknin}) and $s(\cdot)$ is the binary response function of the detectors, which for the sake of simplicity we define here as 

\begin{eqnarray}
s({\it l})= \left\{
\begin{array}{ccccccc}
+1, \ \ \mbox{if} \ \ {\it l} \in [0, +\pi), \\
-1, \ \ \ \mbox{if} \ \ {\it l} \in [-\pi, 0).
\end{array}
\right.
\end{eqnarray}

In order to complete our statistical model we need to specify also the (density of) probability $\rho({\it l})$ of each hidden configuration ${\it l} \in {\cal S}$ over the space ${\cal S}$ to occur in every single realization of the pair of entangled particles. By symmetry considerations this density of probability must be functionally identical from the point of view of both detectors, independently of their relative orientation. Moreover, the condition of 'free-will' demands that the probability of each hidden configuration to occur in any single realization of the experiment cannot depend on the parameterizations of the space ${\cal S}$ associated to each one of the two detectors. This condition can be precisely stated as: 

\begin{equation}
\label{free-will}
d\lambda_A \ \rho(\lambda_A) = d\lambda_B \ \rho(\lambda_B).
\end{equation} 
It is straightforward to show from (\ref{free}) that this condition is fulfilled if and only if the probability density $\rho({\it l})$ is given by:

\begin{equation}
\label{density}
\rho({\it l})=\frac{1}{4}\left|\sin({\it l})\right|.
\end{equation}

We can now compute within the framework of this model the statistical correlations expected  between the outcomes of the two detectors as a function of their relative orientation. The binary outcomes of each one of the two detectors define a partition of the phase space ${\cal S}$ of all the possible hidden configurations into four coarse subsets,
\begin{eqnarray*}
\label{four_subsets}
(s^{(A)}=+1; s^{(B)}=+1) & \Longleftrightarrow & \lambda_A \in [0, \Delta-\Phi) \\
(s^{(A)}=+1; s^{(B)}=-1) & \Longleftrightarrow & \lambda_A \in [\Delta-\Phi, \pi) \\
(s^{(A)}=-1; s^{(B)}=+1) & \Longleftrightarrow & \lambda_A \in [\Delta-\Phi-\pi, 0) \\
(s^{(A)}=-1; s^{(B)}=-1) & \Longleftrightarrow & \lambda_A \in [-\pi, \Delta-\Phi-\pi),
\end{eqnarray*}
where we have assumed without any loss of generality that $\Delta - \Phi \in [0, \pi)$. Each one of these four coarse subsets happen with a probability given by: 
\begin{eqnarray*}
\begin{array}{cccccc}
p\left(+1,+1\right) & =  & \int_0^{\Delta-\Phi} \rho(\lambda_A) \ d\lambda_A \hspace{0.15in} & = \ \frac{1}{4}\left(1 - \cos (\Delta-\Phi)\right), \vspace{0.1in} \\ 
p\left(+1,-1\right)  & =  & \int_{\Delta-\Phi}^{\pi} \rho(\lambda_A) \ d\lambda_A \hspace{0.12in} & = \ \frac{1}{4}\left(1 + \cos (\Delta-\Phi)\right), \vspace{0.1in} \\
p\left(-1, +1\right) & = & \int_{\Delta-\Phi-\pi}^{0} \rho(\lambda_A) \ d\lambda_A & = \ \frac{1}{4}\left(1 + \cos (\Delta-\Phi)\right), \vspace{0.1in} \\
p\left(-1,-1\right) & = & \int_{-\pi}^{\Delta-\Phi-\pi} \rho(\lambda_A) \ d\lambda_A & = \ \frac{1}{4}\left(1 - \cos (\Delta-\Phi)\right).
\end{array}
\end{eqnarray*}
These conditional probabilities reproduce the predictions of quantum mechanics (\ref{correlation}):

\begin{eqnarray}
\nonumber
E(\Delta,\Phi) = p\left(+1,+1\right) + p\left(-1,-1\right) - p\left(+1,-1\right) - p\left(-1,+1\right) = -\cos(\Delta-\Phi). 
\end{eqnarray}

Finally, we notice that in spite of the non-trivial transformation law (\ref{Oaknin}) our model complies with the trivial demand that a relative rotation of the measurement apparatus by an angle $\Delta$ followed by a second relative rotation by an angle $\Delta'$ results into a final rotation by an angle $\Delta+ \Delta'$. 
Consider, for example, an initial reference setting ${\cal T}_0$ in which the outcomes of the two measurement apparatus are correlated by an amount $E=-\cos(\Phi)$. The angular coordinates of the hidden configurations with respect to each one of the two measurement devices, $\lambda_A$ and $\lambda_B$, would be related in this reference setting by the relationship:

\vspace{-0.08in}
\begin{eqnarray}
\label{L0}
\lambda_B =-L(\lambda_A; -\Phi).
\end{eqnarray} 

We now define a new measurement setting ${\cal T}_1$ obtained from the initial setting ${\cal T}_0$ by rotating the relative orientation of the two apparatus by an angle $\Delta$. The angular coordinates $\lambda_A$ and $\lambda'_B$ defined with respect to this new setting would be related by:

\vspace{-0.05in}
\begin{equation}
\label{L1}
\lambda'_B =-L(\lambda_A; \Delta - \Phi).
\end{equation} 
A third measurement setting ${\cal T}_2$ is obtained from the intermediate setting ${\cal T}_1$ by rotating the relative orientation of the two apparatus by an additional angle $\Delta'$. In the intermediate setting ${\cal T}_1$, which is now taken as reference to define the second rotation, the pair of particles appears to be in a polarization state characterized by a phase $\Phi' = -\Delta + \Phi$. Hence, the angular coordinates $\lambda_A$ and $\lambda''_B$ defined with respect to the setting ${\cal T}_2$ would be related by the transformation law:
\vspace{-0.02in}
\begin{equation}
\label{L2}
\lambda''_B =-L(\lambda_A; \Delta' - \Phi') = -L(\lambda_A; \Delta' + \Delta - \Phi).
\end{equation} 
By comparison of the transformation law (\ref{L0}) for the initial setting ${\cal T}_0$ and the transformation law (\ref{L2}) for the setting ${\cal T}_2$, we realize that the latter has been obtained from the initial setting by rotating the apparatus by an angle $\Delta' + \Delta$, as we had demanded.
\

In order to complete the description of the Bell experiment we define two new settings ${\cal T}_3$ and ${\cal T}_4$, which are obtained, respectively, from ${\cal T}_1$ and ${\cal T}_2$ by cancelling the phase $\Phi$ in the reference setting ${\cal T}_0$. Hence, in these settings the angular coordinates of the hidden configurations with respect to the two measurement apparatus are related by the relationships:

\begin{equation}
\label{L3}
\lambda'''_B =-L(\lambda_A; \Delta).
\end{equation} 
and
\begin{equation}
\label{L4}
\lambda''''_B = -L(\lambda_A; \Delta' + \Delta),
\end{equation} 
respectively. Thus, we could intuitively think about the four settings of the detectors involved in a Bell experiment as corresponding to two possible values for the relative angle $\Delta$ and two possible values for the phase $\Phi$, while they all four share the orientation of one the two detectors, say detector A, taken as reference. 

Finally, let us notice that when we substitute the coherent source of pairs of entangled particles (\ref{Bell_state}) by the incoherent {\it classical} source (where all the mixed coherent sources are defined with respect to the same arbitrary setting of the two detectors):

\begin{eqnarray}
{\hat \mu} = \int_{2\pi} \ d\Phi \  |\Psi_{\Phi}\rangle \langle \Psi_{\Phi}| = |\uparrow\rangle \langle \uparrow|^{(A)} \otimes |\downarrow\rangle \langle \downarrow|^{(B)} +  |\downarrow\rangle \langle \downarrow|^{(A)} \otimes |\uparrow\rangle \langle \uparrow|^{(B)},
\end{eqnarray}
the broken rotational symmetries are statistically restored and the outcomes of the two measurement devices become uncorrelated for all settings. Only then, when the rotational symmetries are restored, we can safely define separately the orientations of each one of the measurement devices with respect to some external reference frame and, thus, describe the phase space of its possible settings with the help of these two angles $\left(\Omega_A,\Omega_B\right)$. 
\\
\section{A proposal for an experimental test}
The statistical model of hidden configurations described in the previous section reproduces the quantum mechanical prediction for the correlation (\ref{correlation}) between the binary outcomes of projective polarization measurements performed on each one of the two particles of every entangled pair, as a function of the angular parameter $\Delta-\Phi$ that characterizes the experimental setting. However, with the help of additional weak polarization mesurements the predictions of this statistical model can still be experimentally distinguished from those of the standard framework of quantum mechanics.

Let us consider as before a source of pairs of entangled particles prepared in a Bell state (\ref{Bell_state}) and a pair of measuring devices that test their polarizations through projective measurements at a relative angle $\Delta-\Phi=\pi/4$, so that the correlation between their binary outcomes is $E_{A_1,B_2}=E(\pi/4)=-1/\sqrt{2}$. For reasons that will be immediately clear we denote this correlation as $E_{A_1,B_2}$. This correlation is only very slightly modified if we perform on particle B a very weak polarization measurement before the projective polarization test \cite{Korotkov,Dressel}. If we design the weak measurement on particle B so that it is oriented along a relative angle $\Delta-\Phi=-\pi/4$ with respect to the projective polarization measurement on particle A, the correlation between their outcomes in a long sequence of repetitions will be given by $E_{A_1,B_1}=E(-\pi/4)=-1/\sqrt{2}$.

We can now ask ourselves what would be the correlation $E_{B1,B2}$ between the outcomes of the weak measurement performed on particle B and the projective measurement performed on the same particle later on. According to quantum mechanics their correlation should be
\begin{equation}
E^{QM}_{B1,B2} = \cos(\pi/2) = 0,
\end{equation}
while in the statistical model presented in the previous section their correlation would be \cite{david4}
\begin{eqnarray}
\nonumber
E^{SM}_{B1,B2} = 4 \left(\int_{\pi/4}^{\pi/2} \ \rho(\lambda) \ d\lambda \ - \ \int_0^{\pi/4} \ \rho(\lambda) \ d\lambda\right)
= \hspace{2.0in}\\
\nonumber
= \int_{\pi/4}^{\pi/2} \ \left|\sin(\lambda)\right| \ d\lambda \ - \int_0^{\pi/4} \ \left|\sin(\lambda)\right| \ d\lambda = \hspace{1.65in}\\
\nonumber
=-\cos(0) + \cos(\pi/4) - \cos(\pi/2) + \cos(\pi/4) = \hspace{0.9in} \\ 
= \sqrt{2} - 1 \simeq 0.41 \neq E^{QM}_{B1,B2}. \hspace{0.95in}
\end{eqnarray}
\section{Discussion}

The Bell theorem is one of the pillars upon which relies the widely accepted belief that quantum mechanics is the ultimate mathematical framework within which the hypothetical final theory of the fundamental building blocks of Nature and their interactions must be formulated. The theorem proves through an experimentally testable inequality (the Bell inequality) that the predictions of quantum mechanics for the Bell polarization states of two entangled particles cannot be reproduced by any underlying theory of hidden variables that shares certain intuitive features.

In this paper we have shown, however, that these intuitive features include a subtle, though crucial, assumption that is not required by fundamental physical principles and, hence, it is not necessarily fulfilled in the actual experimental setup that tests the inequality. In fact, the disputed assumption cannot be implemented within the framework of standard quantum mechanics either.

Namely, the proof of the Bell theorem requires the existence of a preferred frame of reference, supposedly provided by a lab, with respect to which the orientations of each one of the two measurement devices can be independently defined for every single realization of the experiment. This preferred frame is required in order to compare the orientations of the detectors in a sequence of repetitions of the experiment, since in every realization each particle's polarization can be tested along a single orientation. 

Notwithstanding, the existence of a preferred frame of reference is at odds with Galileo's fundamental principle of relativity and, indeed, it cannot exist when the hidden configurations of the pair of entangled particles spontaneously break the rotational symmetry of the experimental setup under rigid rotations of the two detectors and a non-zero geometric phase accumulates through cyclic gauge transformations. In such a case, in order to compare different realizations of the experiment, we must pick the orientation of one of the detectors as a common reference direction, with respect to which the relative orientation of the second detector is defined. Under these conditions the Bell theorem does not necessarily hold, see (\ref{new_model}),  Fig. 2 and Fig. 3.  

Following these ideas we explicitly built a model of hidden variables for the Bell states of two entangled particles that reproduces the predictions of quantum mechanics. Further details of the model are discussed in \cite{david}. In two additional accompanying papers we have used these same ideas to build explicit local models of hidden variables for the GHZ state of three entangled particles \cite{david2} and also for the qutrit \cite{david3}. 

The derivation of a model of local hidden variables for the entangled states of two or more qubits means that entanglement, the quintessential quantum phenomenon, can be fully described without the quantum formalism. Indeed, the model shows that entanglement can be described in terms of classical statistical concepts, with the help of the well-understood classical notions of curved spaces and gauge degrees of freedom. Thus, the model proves that there are not mysterious fundamental differences between classical and quantum correlations.

Furthermore, the model of hidden variables presented here opens the window to the possible existence of an unexplored physical reality that might underlay the laws of quantum mechanics \cite{Ball} and, thus, it might lead to a whole new area of research in physics in quest for the fundamental laws of this underlying reality. The existence of such a reality was first suggested 85 years ago by Einstein, Podolsky and Rosen through their famous EPR paradox \cite{EPR,Bohm}, but following Bell's arguments it had been thought that an underlying reality was incompatible with quantum mechanics \cite{Bell2,K-S}.

Finally, we wish to notice that our model of hidden variables is built upon fundamental physical concepts shared by the formalism of General Relativity and, thus, it might eventually lead to a unified description of quantum phenomena and gravitation.

\section{Acknowledgements}

This manuscript has been released as a pre-print at https://arxiv.org/abs/1912.06349, D.H.~Oaknin \cite{david5}.
Its author D.H.~Oaknin was employed by the company RAFAEL Advanced Defense Systems.


\begin{references}

\bibitem{Bell} J.S.~Bell, "On the Einstein-Podolsky-Rosen paradox", Physics {\bf 1964}, 1, 195-200.

\bibitem{Fine} A.~Fine, "Hidden variables, joint probability, and the Bell inequalities", Phys. Rev. Lett. {\bf 1982}, 48, 291.

\bibitem{Hansen} B.~Hensen {\it et al}, "Loophole-free Bell inequality violation using electron spins separated by 1.3 kilometres", Nature {\bf 2015}, 526, 682.

\bibitem{Aspect} A.~Aspect, J.Dalibard and G.~Roger, "Experimental test of Bell's inequalities using time-varying analyzers", Phys. Rev. Lett. {\bf 1982}, 49, 1804.

\bibitem{Gisin} W.~Tittel, J.~Brendel, H.~Zbinden and N.~Gisin, "Violation of Bell Inequalities by Photons More Than 10 km Apart", Phys. Rev. Lett. {\bf 1998}, 81, 3563.

\bibitem{Weihs} G.~Weihs et al., "Violation of Bell's inequality under strict Einstein locality conditions", Phys. Rev. Lett. {\bf 1998}, 81, 5039. 

\bibitem{Rowe}  M.A.~Rowe, D.~Kielpinski, V.~Meyer, C.A.~Sackett, W.M.~Itano, C.~Monroe, D.J.~Wineland, "Experimental violation of a Bell's inequality with efficient detection", Nature {\bf 2001}, 409, 791.

\bibitem{Giustina} M.~Giustina, A.~Mech, S.~Ramelow, B.~Wittmann, J.~Kofler, J.~Beyer, A.~Lita, B.~Calkins, T.~Gerrits, S.W.~ Nam, R.~ Ursin and A.~Zeilinger, "Bell violation using entangled photons without the fair-sampling assumption", Nature, {\bf 2013}, 497, 227.

\bibitem{Christensen} B.G.~Christensen, K.T.~McCusker, J.~Altepeter, B.~Calkins, T.~Gerrits, A.~Lita, A.~Miller, L.K.~Shalm, Y.~Zhang, S.W.~Nam, N.~Brunner, C.C.W.~Lim, N.~Gisin and P.G.~Kwiat, "Detection-Loophole-Free Test of Quantum Nonlocality, and Applications". Phys. Rev. Lett. {\bf 2013}, 111, 130406. 

\bibitem{Giustina2} M.~Giustina, M.A.M.~Versteegh, S.~Wengerowsky, J.~Handsteiner, A.~Hochrainer, K.~Phelan, F.~Steinlechner, J.~Kofler, J.~Larsson, C.~Abellan, W.~Amaya, V.~ Pruneri, M.W.~Mitchell, J.~Beyer, T.~Gerrits, A.E.~Lita, L.K.~Shalm, S.W.~Nam, T.~Scheidl, R.~Ursin, B.~Wittmann and A.~Zeilinger, "A significant-loophole-free test of Bell's theorem with entangled photons". Phys. Rev. Lett. {\bf 2015}, 115, 250401. 

\bibitem{Shalm} L.K.~Shalm, E.~Meyer-Scott, B.G.~Christensen, P.~Bierhorst, M.A.~Wayne, M.J.~Stevens, T.~Gerrits, S.~ Glancy, D.R.~Hamel, M.S.~Allman, K.J.~Coakley, S.D.~Dyer, C.~Hodge, A.E.~Lita, V.B.~Verma, C.~Lambrocco, E.~Tortorici, A.~Migdall, Y.~Zhang, D.R.~Kumor, W.H.~Farr, F.~Marsili, M.~Shaw, J.A.~Stern, C.~Abellán, W.~Amaya, V.~Pruneri, T.~Jennewein, M.W.~Mitchell, P.G.~Kwiat, J.C.~Bienfang, R.P.~Mirin, E.~Knill, S.W.~Nam, "A strong loophole-free test of local realism", Phys. Rev. Lett. {\bf 2015}, 115, 250402.

\bibitem{ScienceNews} H.~Wiseman, "Quantum physics: Death by experiment for local realism", Nature {\bf 2015}, 526, 649.

\bibitem{CHSH} J.F.~Clauser, M.A.~Horne, A.~Shimony and R.A.~Holt, "Proposed experiment to test local hidden variables theories", Phys. Rev. Lett. {\bf 1969}, 23, 880–884. DOI: 10.1103/PhysRevLett.23.880.

\bibitem{Tsirelson} B. S.~Cirelson, "Quantum generalizations of Bell's inequality", Lett. Math. Phys. {\bf 1980}, 4, 93.

\bibitem{Hess} K.~Hess, "Kolmogorov's probability spaces for 'entangled' data subsets of EPRB experiments: no violation of Einstein's separation principle", Journal of Modern Physics {\bf 2020}, 11, 5. 

\bibitem{Hess1} K.~Hess, "Bell's theorem and Instantaneous influences at a distance", arXiv:1805.04797.

\bibitem{Hess2} K.~Hess and W.~Philipp, "Bell's theorem and the problem of decidability between the views of Einstein and Bohr",
PNAS {\bf 2001} 98 (25) 14228.

\bibitem{Hess3} K.~Hess and W.~Philipp, "Breakdown of Bell's theorem for certain objective local parameter spaces",
PNAS {\bf 2004} 101 (7) 1799.

\bibitem{Hess4} K.~Hess, H.~De Raedt and K.~Michielsen, "From Boole to Leggett-Garg: Epistemology of
Bell-Type Inequalities", Advances in Mathematical Physics, ID 4623040 (2016).

\bibitem{Wilczek}  F.~Wilczek and A.~Shapere, eds. (1989). Geometric Phases in Physics. Singapore: World Scientific.

\bibitem{PRboxes} S.~Popescu and D.~Rohrlich, "Quantum nonlocality as an axiom", Foundations of Physics {\bf 1994}, 24, 379.

\bibitem{Revzen} M.~Revzen, "Kolmogorov proof of the Clauser-Horne-Shimony-Holt inequalities", Int. Journal of Quantum Information, {\bf 16},  2,  1850013 (2018).

\bibitem{Korotkov} T.C.~White, J.Y.~Mutus, J.~Dressel {\it et al}, "Preserving entanglement during weak measurement demonstrated with a violation of the Bell-Leggett-Garg inequality", npj Quantum Information {\bf 2016}, 2, 15022.

\bibitem{Dressel} J.~Dressel and A.N.~Korotkov, "Avoiding loopholes with hybrid Bell-Leggett-Garg inequalities", Phys. Rev. A {\bf 2014}, 89, 012125.

\bibitem{Ball} P.~Ball, "Exorcising Einstein's spooks", Nature (2001). doi:10.1038/news011129-15.

\bibitem{EPR} A.~Einstein, B.~Podolsky and N.~Rosen, "Can quantum mechanical description of physical reality be considered complete ?",  Phys. Rev. {\bf 1935}, 47, 777-780, DOI: 10.1103/Phys.Rev.47.777.

\bibitem{Bohm} D.~Bohm,  Quantum Theory {\bf 1951}, Prentice-Hall, New York.

\bibitem{Bell2} J.S.~Bell, "On the problem of hidden variables in quantum mechanics", Physics {\bf 1966}, 38, 447-452.

\bibitem{K-S} S.~Kochen and E.P.~Specker, "The problem of hidden variables in quantum mechanics", J. Math. Mech. {\bf 1967}, 17, 59-87.

\bibitem{david} D.H.~Oaknin, "Solving the EPR paradox: an explicit statistical local model of hidden variables for the singlet state", arXiv:1411.5704.

\bibitem{david2} D.H.~Oaknin, "Solving the Greenberger-Horne-Zeilinger paradox: an explicit local model of hidden variables for the GHZ state", arXiv:1709.00167.

\bibitem{david3} D.H.~Oaknin, "Bypassing the Kochen-Specker theorem: and explicit non-contextual model of hidden variables for the qutrit", arXiv:1805.04935. 

\bibitem{david4} D.H.~Oaknin, "Comment on 'White, T., Mutus, J., Dressel, J. et al., 'Preserving entanglement during weak measurement demonstrated with a violation of the Bell-Leggett-Garg inequality', npj Quantum Information 2, 15022 (2016)",  
hal-02554478.

\bibitem{david5} D.H.~Oaknin, "The Bell theorem revisited: geometric phases in gauge theories", arXiv:1912.06349.
\end{references}
\end{document}